\documentclass[superscriptaddress,aps,preprint,amsmath,amssymb,floatfix]{revtex4-1}

\usepackage{graphicx,morefloats}
\usepackage{subfigure} 
\usepackage{color,soul}
\usepackage{bm}
\usepackage{amsmath}
\usepackage{amssymb}
\usepackage{times}
\usepackage{hyperref}
\usepackage{bbm}

\begin{document}

\hyphenation{va-ni-sh-ing}

\begin{center}

\thispagestyle{empty}

{\large\bf Fragile Insulator and Electronic Nematicity in a Graphene  Moir\'{e} System}
\\[0.6cm]

Lei\ Chen$^{\dagger}$, Haoyu\ Hu$^{\dagger}$, Qimiao\ Si$^{\ast}$ \\[0.0cm]

Department of Physics and Astronomy \& Rice Center for Quantum Materials, Rice University, Houston, Texas 77005, USA\\[-0.cm]

\end{center}

\vspace{0.3cm}

{\bf Strongly correlated quantum matter
exhibits a rich variety of remarkable properties,  but the organizing principles 
that underlie the behavior remain to be established. Graphene heterostructures, which can host narrow 
moir\'{e} electron bands\cite{Bistritzer2011} that amplify the correlation effect, represent a new setting to make progress 
on this overarching issue. In  such correlated moir\'{e} systems, an insulating state is a prominent feature
of the phase diagram and may hold the key to understanding the  basic physics.
Here we advance the notion of a fragile insulator, a correlation-driven insulating state that 
is on the verge of a delocalization transition into a bad metal. Using a realistic multiorbital Hubbard model
as a prototype for narrow band moir\'{e} systems, we realize such a fragile insulator and demonstrate a 
nematic order in this state as well as in the nearby bad metal regime. Our results are consistent with
the observed electronic anisotropy in the graphene moir\'{e} systems\cite{Kerelsky2019,Choi2019,Jiang2019,Cao2020nematicity}
and provide a natural understanding of what happens when the insulator is tuned into a bad 
metal\cite{Stepanov2020Metallic,Saito2020Metallic,Arora2020Metallic}. We propose the fragile insulator 
and the accompanying bad metal as competing states at integer fillings that analogously anchor
the overall phase diagram of the correlated moir\'{e} systems and beyond.
}
\vspace{0.6cm}

\noindent E-mail: $^{\ast}$qmsi@rice.edu

\noindent $^{\dagger}$L.\,C. and H.\,H. contributed equally to this work.

\newpage
Strongly correlated systems are epitomized by cuprate superconductors, where a robust insulating state
serves as the 
``parent"
 from which high temperature superconductivity develops 
upon introducing charge carriers\cite{Lee_RMP2006}.
Correlated insulators and superconductivity also
emerge
in the twisted bilayer graphene (TBG) at magic angles\cite{Cao2018Magic,Cao2018correlated,Lu2019,Yankowitz2019},
trilayer graphene heterostructures with hexagonal boron nitride substrate (TLG/hBN)\cite{Chen2019-1,Chen2019-2}
and related structures. 
The narrowness of the moir\'{e} bands at the magic angles\cite{Bistritzer2011}
implies that the relative strength of the electron correlations is enhanced, 
which has been
demonstrated by spectroscopic means\cite{Kerelsky2019,Choi2019,Jiang2019,Xie2019}.
 The insulators appear at the partial but integer fillings of the moir\'{e} bands, 
 where the system would have been metallic in the absence of electron correlations,
 while superconductivity arises when 
the charge carrier concentration is tuned away from such fillings.
By analogy with the cuprates,
the insulating phase is believed to be key to elucidating the correlation physics of the moir\'{e}
systems\cite{Xu2018,Dodaro2018,Padhi2018,Thomson2018,Xie2020,Pizarro2019}.
However, understanding the insulators
remains a pressing open question.

Here, we address this issue, departing from several motivating factors. 
One consideration concerns the insulating nature {\it per se}. The insulating behavior
develops at energy scales that can be low compared to either the effective Coulomb repulsion 
($U$) or the width ($W$) of the moir\'{e} bands. For instance, for the magic-angle TBG devices
at half filling ($\nu=2$, corresponding to $1$ electron or hole per valley per moir\'{e} unit cell), the electrical resistivity shows an 
insulating-like temperature dependence below about $4$ K (Refs.\,\citenum{Cao2018Magic,Cao2018correlated,Dodaro2018}),
which is more than one decade lower than the scale $U \sim W \sim 10$ meV. Moreover, recent experiments have shown that
the insulator can be tuned away quantum mechanically:
This happens upon varying the strength of the electron correlations without changing the carrier concentration,
while superconductivity persists; the result raises  the question of whether the insulator anchors the phase diagram 
at all\cite{Stepanov2020Metallic,Saito2020Metallic,Arora2020Metallic}.

Another consideration is about electronic orders, a rich landscape of which is 
one of the salient characteristics of strongly correlated  electron systems\cite{Lee_RMP2006,Si16.1,Kei17.1}.
Recently, measurements by scanning tunneling microscopy (STM)\cite{Kerelsky2019,Choi2019,Jiang2019}
have revealed evidence for electronic nematic correlation
in the normal state (above the superconducting transition temperature) of the magic-angle TBG.
The local density of states shows a three-fold anisotropy\cite{Kerelsky2019}, implying a large nematic susceptibility and 
possibly even a nematic order. Importantly, the effect maximizes near the insulating 
phase of the half-filled moir\'{e} bands ({\it i.e.}, two electrons or holes per unit cell of the moir\'{e} superlattice)\cite{Kerelsky2019}.
These STM observations are complemented by transport measurements\cite{Cao2020nematicity},
which furthermore connect the nematicity with superconductivity.
Understanding the nematic correlation is important, as it is primed for clues about the underlying correlation physics.

\begin{figure}[t]
\includegraphics[width=0.9\columnwidth]{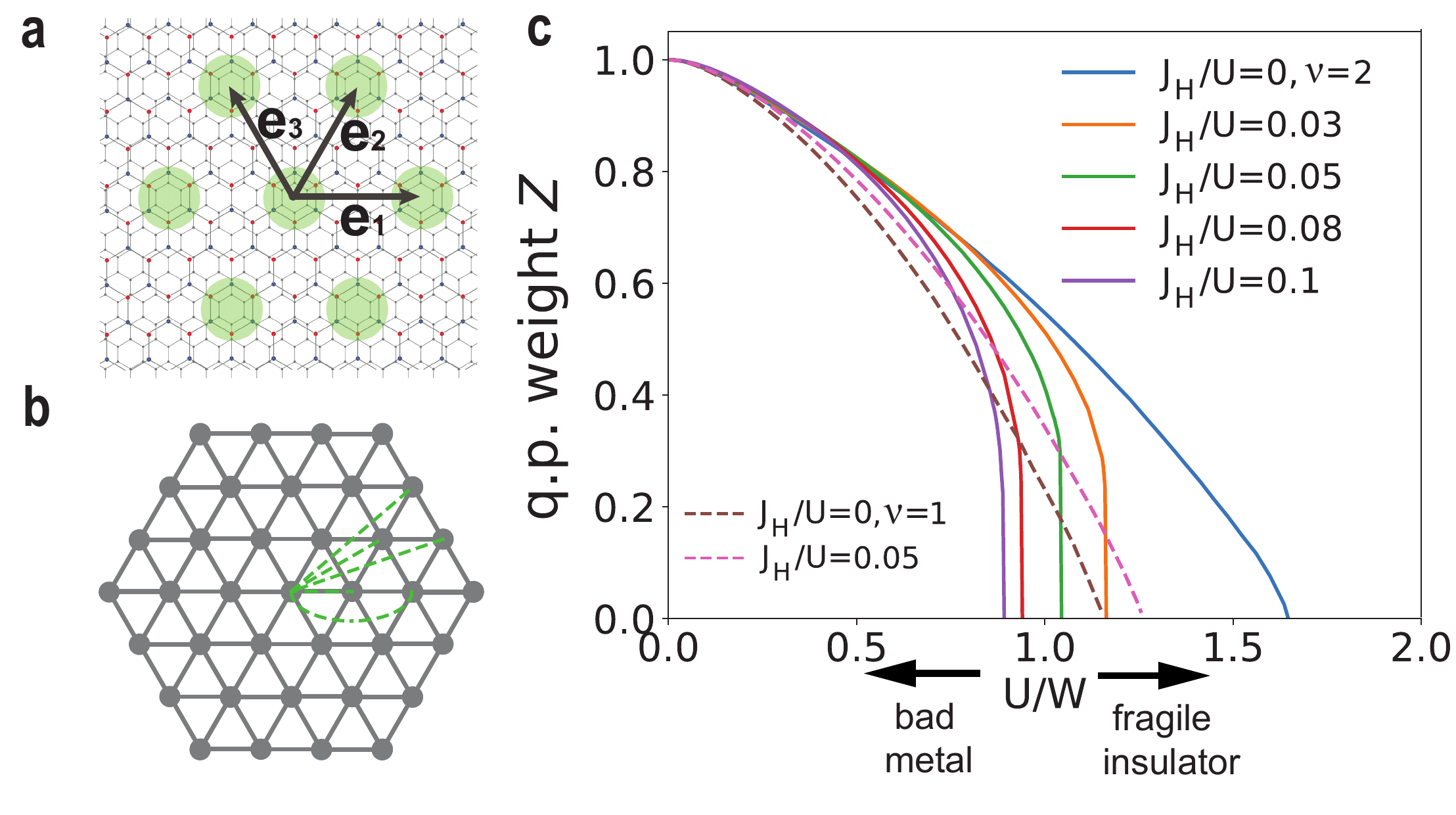}
\vskip 0.15 in
\caption{ {\bf A graphene moir\'{e} system and the development of fragile insulator and bad metal.}
{\bf a,} Illustration of the mori\'e superlattice, where ${\bf e_1}, {\bf e_2}, {\bf e_3}$ denote the superlattice basis
vectors. The triangular lattice, marked by the green regions, results from a difference in the lattice constants between 
TGL and hBN. 
  {\bf b,} The bonds (dashed lines) for the 
hopping parameters, $t_1-t_5$, of the effective tight binding model,
which  specify the hopping parameters for the other bonds in the moir\'{e} superlattice 
through the $C_6$ and $M_y$ transformations
(see Methods and Supplementary Information).
{\bf c,} The quasiparticle (q.p.) weight $Z$ as a function of $U/W$. The results show the strong influence of the Hund's coupling 
$J_H$ on the metal-insulator transition.
 For $J_H/U=0,0.03,0.05,0.08,0.1$ at the half filling
 ($\nu=2$) of the moir\'{e} bands, the Mott transition thresholds are $U_c/W=1.65, 1.16, 1.04,0.94,0.89$, respectively.
 For $J_H/U=0, 0.05$ at the quarter filling ($\nu=1$), they are  $U_c/W=1.15,1.26$, respectively.
 }
\label{qpz}
\end{figure}

The system we choose to focus on as a prototype case is TLG/hBN,  where
the correlation physics can be isolated 
and non-perturbative theoretical analyses are possible.
In this system,  the moir\'{e} superstructure (Fig.\,\ref{qpz}a)
results from a small difference 
between the in-plane lattice constants of the ABC stacked TLG (see Supplementary Information, Fig.\,\ref{abc}) and
hBN\cite{Chen2019-1,Chen2019-2,Zhu2018,Zhang2019Bridging}. 
In the case of the magic-angle TBG, a topological obstruction to the construction of Wannier orbitals for their moir\'{e} bands 
has been actively discussed (for example, Refs.\,\citenum{Po2018,NFQYuan2018}). 
The TLG/hBN structure under a particular direction of the perpendicular 
voltage bias, while having electronic properties with considerable similarities to those of the magic-angle TBG, do
not 
face such an obstruction\cite{Chen2019-1,Chen2019-2}.
Consequently, their moir\'{e} bands are faithfully represented by a two-orbital 
Hubbard Hamiltonian\cite{Zhang2019Bridging},
comprising the kinetic part, $H_0$, and the interaction part $H_U+H_V$ (see Methods).
The reggime of prime interest corresponds to intermediate correlation,
with the normalized interaction $U/W \sim O(1)$, which is difficult to access
by perturbative expansions either in $U/W$ or its inverse.
 Here, we investigate this regime using non-perturbative methods.
 Our primary tool will be the recently developed Variational Monte Carlo (VMC) method
 that incorporates the correlation effects of not only the Hubbard interaction but also 
 the Hund's coupling\cite{Hu2019}
 (see Methods).

We consider the
half-filled
case,
 keeping in mind the aforementioned motivations.
To be definite, we focus on the case with a perpendicular voltage bias,
which fixes the tight-binding parameters
(see Fig.\,\ref{qpz}b, Methods and Supplementary Information) for $H_0$,
and allows for an estimate of the parameters for 
both $H_U$ (the onsite Hubbard interaction $U$ and Hund's coupling $J_H$)
and
$H_V$ (the density-density interaction $V$ and spin-valley exchange interaction $V_H$ between 
nearest-neighbor 
sites)\cite{Zhang2019Bridging}.
A metal-insulator transition (MIT) could arise in the multiorbital model, in spite of having an even number of electrons 
per unit cell, due to the onsite interactions. We address their effects 
by performing a saddle-point  analysis within a U(1) slave-spin method\cite{Yu2012} (see Methods).
Focusing on the paramagnetic phase that preserves the time reversal and translation symmetries,
the quasiparticle weight is the same for the two valleys,
$Z_{+} = Z_{-}=Z$.
The results of  $Z$ {\it vs.} $U/W$, for various values of the ratio $J_H/U$, are shown in Fig.\,\ref{qpz}c. 
In the absence of the Hund's coupling, $J_H=0$, a metal-to-insulator transition occurs at $ U_c (J_H=0) /W \approx 1.65$.
As Fig.\,\ref{qpz}c also shows, even a relatively small 
Hund's coupling considerably enhances the localization effect, and turns the threshold
for the metal-insulator transition to about $U_c/W \approx 1$.

The metallic regime in proximity of the metal-to-insulator transition corresponds to a bad metal, where the quasiparticle 
weight $Z$ is much  reduced from the free-electron value $1$ (Refs.\,\citenum{Si16.1,Hus04.1}). 
This motivates us to dub the insulating  regime in proximity of the insulator-to-metal transition a fragile insulator. 
In this regime, the insulating gap is considerably smaller than either $U$ and $W$. 
Correspondingly, the temperature scale for the onset of the insulating 
behavior is expected to be small compared to $U/k_B$ and $W/k_B$.
We expect such behavior to occur in other integer fillings between charge neutrality and fully
filled moir\'{e} bands, {\it albeit} with a different threshold interaction for the transition. For instance, a similar transition from bad 
metal to fragile insulator occurs 
at quarter filling ($\nu=1$), as is also shown in Fig.\,\ref{qpz}c.

\begin{figure}[t]
\includegraphics[width=0.85\columnwidth]{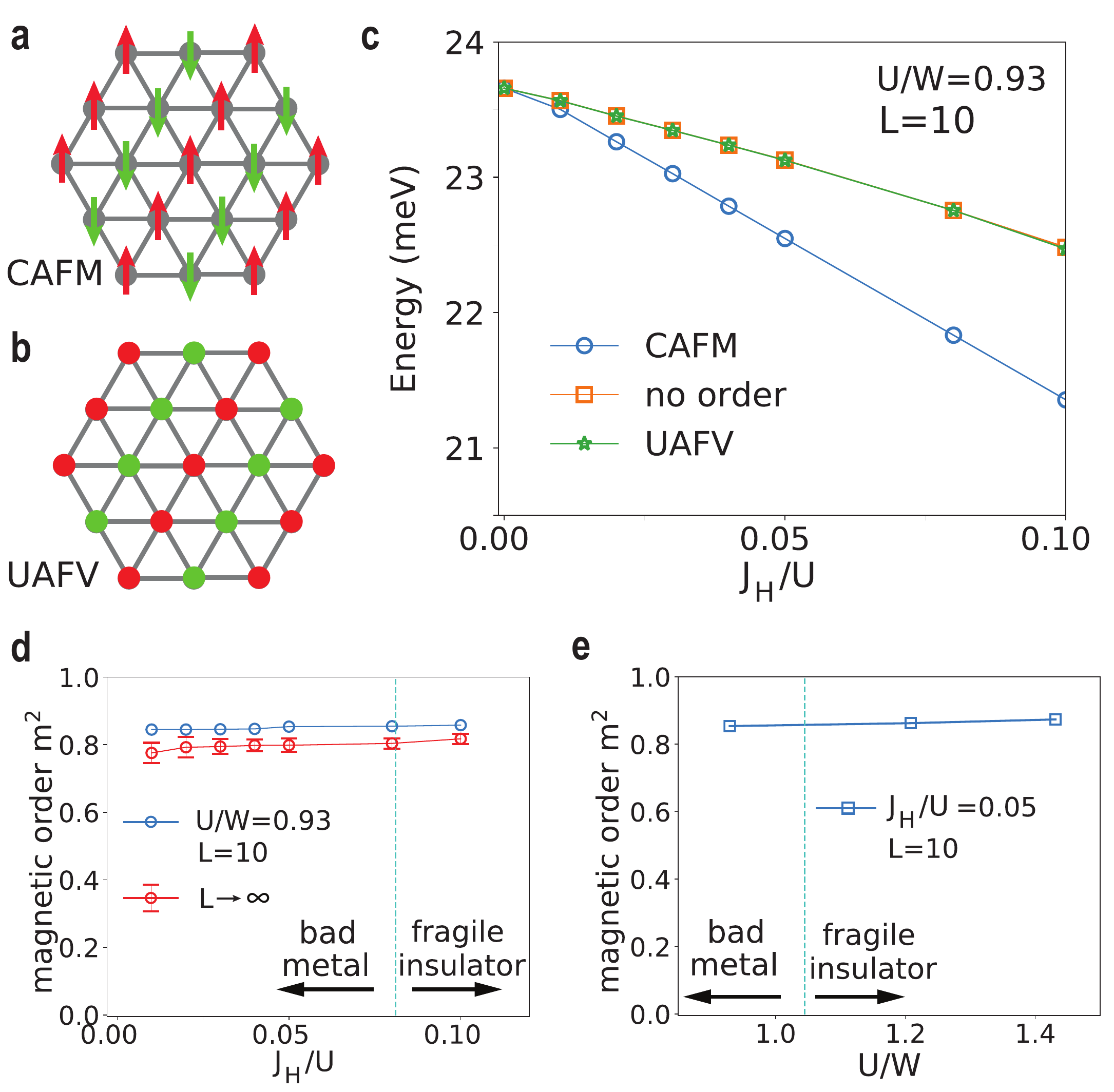}
\vskip 0.15in
\caption{{\bf Ground state at half filling.}
{\bf a,} Illustration of the collinear antiferromagnetic order (CAFM). The red and green spins are opposite of each other.
{\bf b,} That of the uniaxial antiferrovalley order (UAFV). The red (green) sites represent $(n_{+}-n_{-})$ $>$($<$) $0$. 
{\bf c,} The ground state energy of the different states as a function of $J_H/U$ for a fixed $U/W$. The VMC calculation 
is performed on a $L \times L$ geometry (Supplementary Information, Fig.\,\ref{df}).
{\bf d,} The magnetic order parameter ($m^2$)  as a function of the Hund's coupling $J_H/U$ for 
a fixed Hubbard interaction $U/W=0.93$,
at $L=10$ and estimated from a finite size scaling
($L \rightarrow \infty$; see Supplementary Information,
Fig.\,\ref{ord}c,d).
The magnetic order persists for $J_H/U $ as small as $0.01$. 
{\bf e,} The magnetic order parameter ($m^2$) as a function 
of $U/W$ for fixed $J_H/U=0.05$.
}
\label{en}
\end{figure}

We next analyse the possible electronic orders. The effect of onsite interactions is studied using the  VMC method
that is non-perturbative in both the Hubbard interaction and Hund's coupling\cite{Hu2019} (see Methods).
 In the intermediate correlation regime, we find that the ground state has a collinear antiferromagnetic (CAFM) order 
 (Fig.\,\ref{en}a), with the pitch wavevector located at
 ${\bf{Q}}={\bf M}
 =(0,\frac{2\sqrt{3}}{3}\pi)$ of the moir\'{e} Brillouin zone (BZ)
[or equivalently $(\pi,\frac{\sqrt{3}}{3}\pi)$ and $(-\pi,\frac{\sqrt{3}}{3}\pi)$)]. Fig.\,\ref{en}c 
shows  its energy to be lower than that of not only the 
paramagnetic phase ({\it i.e.}, without any order) but also  the competing uniaxial antiferrovalley (UAFV) order (Fig.\,\ref{en}b).
The corresponding magnetic order parameter
$m^2 $ (see Methods) is shown as a function of 
$J_H/U$ for a fixed $U/W$ (Fig.\,\ref{en}d) and {\it vs.} $U/W$ for a fixed $J_H/U$
(Fig.\,\ref{en}e)
as a function of $J_H/U$.  The magnetic order exits both in the fragile insulating and bad metal regimes.

\begin{figure}[t]
\includegraphics[width=0.85\columnwidth]{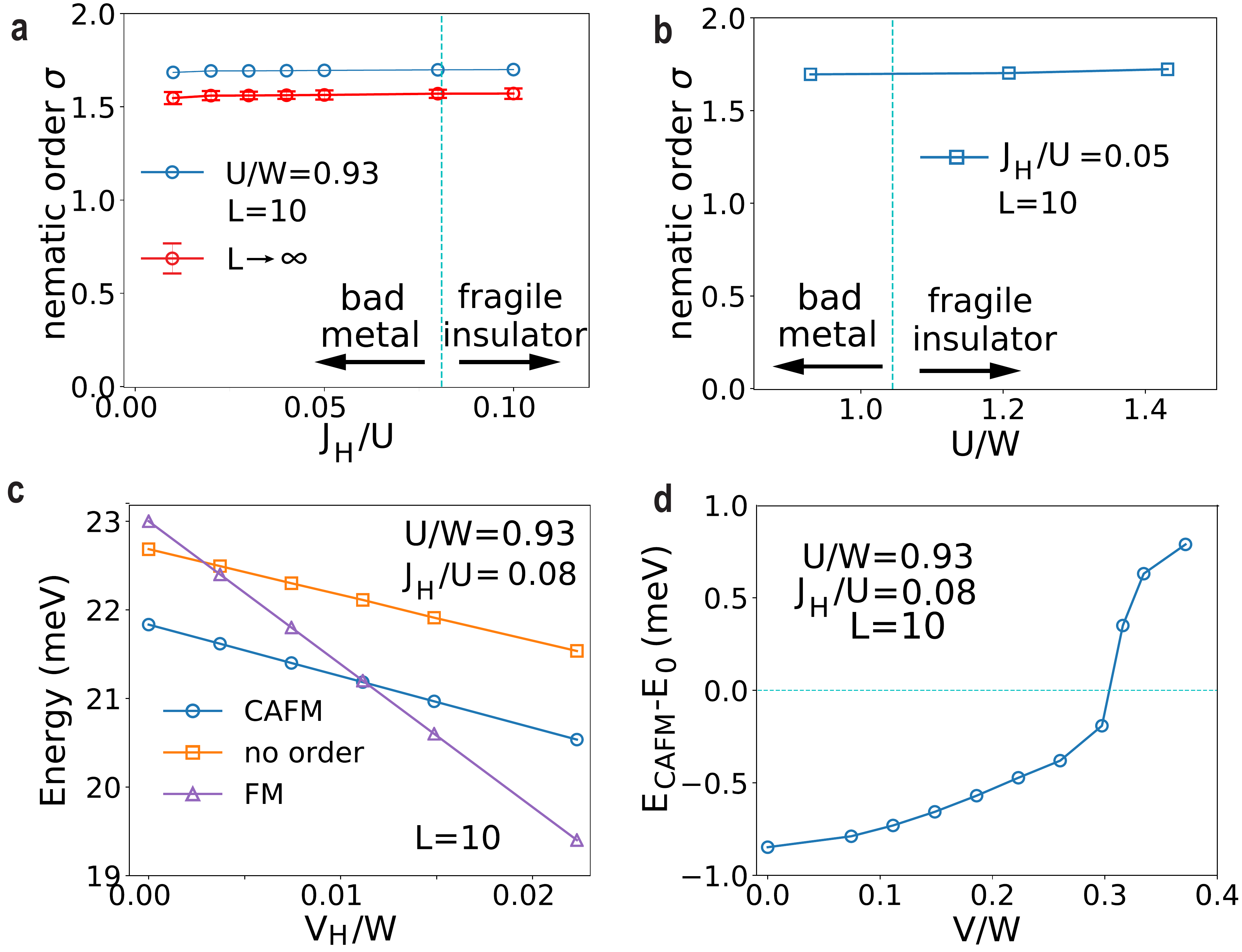}
\vskip 0.15 in
\caption{ 
{\bf Nematic order and its stability.}
{\bf a,} The nematic order parameter ($\sigma$) as a function of the Hund's coupling $J_H/U$ for fixed Hubbard interaction $U/W=0.93$,
calculated at $L=10$ and 
estimated from a finite size scaling
($L \rightarrow \infty$,
see Supplementary Information,
Fig.\,\ref{ord}c,d).
{\bf b,} The nematic order parameter ($\sigma$) as a function  of $U/W$ for fixed $J_H/U=0.05$.
{\bf c,} The energy of the CAFM, non-ordered and ferromagnetic (FM) states {\it vs.} the nearest-neighbor exchange interaction
$V_H/W$ at $J_H/U=0.08$ and $U/W=0.93$. The threshold value for the FM state
to have
a lower energy is 
$V_{H}^1/W \approx 0.011$.
{\bf d,} The difference in the ground state energy between CAFM and the non-ordered state {\it vs.} the nearest-neighbor repulsion
$V/W$ at $J_H/U=0.08$ and $U/W=0.93$. The crossing interaction strength is $V^1/W \approx 0.31$.
}
\label{jhn}
\end{figure}

This sets the stage to  examine the nematicity. The nematic order is classified
  in terms of the breaking of the $C_6$ symmetry, which 
 is an approximate symmetry of the system and exists in the model Hamiltonian, or the $C_3$ symmetry, 
 which survives the weak couplings that exist in the system beyond the 
 model\cite{Chen2019-1,Chen2019-2,Zhu2018,Zhang2019Bridging}.
 The irreducible representations of the  crystal point groups $D_6$ and $D_3$ are given in Table\, \ref{table:symm}.
 We find the relevant nematic order parameter $\sigma$ to be in the $E_2/E$ representation, 
respectively in the $D_6$/$D_3$ classification scheme (see Methods).
The calculated nematic order parameter is shown in Fig.\,\ref{jhn}a,b, respectively 
as a function of $J_H/U$ and {\it vs.} $U/W$.
As a key result of our work, the nematic order parameter is nonzero both in the fragile insulator and bad metal regimes and, 
moreover, it varies smoothly between the two regimes.

It is important to assess the stability of this ground state against the intersite interactions, which are significant due to the 
size of the moir\'{e} unit cell. Consider first the nearest-neighbor  exchange coupling $V_H$.
Our VMC calculation finds the CAFM and associated nematic order 
to be stable for a range of this coupling, up to 
$V_H^{1}/W \approx 0.011$ (Fig.\,\ref{jhn}c).
 Above $V_H^1$, a ferromagnetic order becomes the ground state. 
Consider next the effect of the nearest-neighbor Coulomb repulsion $V$. 
The CAFM and associated nematic order are stable against the paramagnetic phase
for a range of this coupling, up to $V^1/W \approx 0.31$ (Fig.\,\ref{jhn}d).
It is instructive to note that the VMC approach is non-perturbative and, therefore,
advantageous in the intermediate correlation regime [$U/W \sim O(1)$] of interest here.
For comparison,  a self-consistent Hartree-Fock calculation is also carried out. It
 qualitatively captures the transition from CAFM to FM with 
the
increasing 
exchange interaction
$V_H$ (see Supplementary
Information, Fig.\,\ref{mf}a), but 
misses the 
density-density interaction
$V$-induced
 instability of the CAFM phase
 towards the paramagnetic phase
(Supplementary Information, Fig.\,\ref{mf}b). This
result reflects the underestimation of the correlation effect
by the Hartree-Fock method, especially for the paramagnetic state.
Nonetheless, the Hartree-Fock calculation suggests that 
a sufficiently large $V$ makes a charge order viable.
The ordering wavevector is  ${\bf K}$ (Fig.\,\ref{mf}b);
it is a three-sublattice order and is 
not expected to be accompanied by a nematic order.
Note that our purpose is to use the well-defined Hamiltonian as a means to access the 
qualitative features of the overall phase diagram.
Still, the threshold values we have determined,
$V_H^1/W$ and $V^1/W$, are competitive against 
the order-of-magnitude estimates for these parameters
(which, at $\Delta_V=-20$ meV,
are about $0.007$ and $0.37$, respectively\cite{Zhang2019Bridging}),
suggesting that either a nematic order or an enhanced nematic fluctuation is to be expect for TLG/hBN.

We next turn to 
the experimental consequences of our results. First, the proximity of the 
fragile insulator to the delocalization transition implies the development of an energy scale that is 
small compared to the bare energies $U$ and $W$.
The insulating behavior only appears below this scale; above it, the system cannot be distinguished 
from what happens in the bad-metal regime. This provides
 a natural understanding of the experimental observations
that the insulating-like temperature dependence appears in the electrical resistivity 
only at low temperatures \cite{Cao2018Magic,Cao2018correlated}. Second, our finding that the fragile
insulator is nematic allows for a microscopic understanding of the nematic correlations that have 
been observed in the normal state of the graphene 
moir\'{e} systems \cite{Kerelsky2019,Choi2019,Jiang2019,Cao2020nematicity}.
This is especially so given that the electronic anisotropy has been experimentally demonstrated
to be peaked near half filling
\cite{Kerelsky2019}.
Third, we have found the electronic nematic order to appear at half filling, 
not only when the system is a fragile insulator but also
when it is a  bad metal.
This leads us to predict that
 devices where the insulating phase has turned metallic\cite{Stepanov2020Metallic,Saito2020Metallic,Arora2020Metallic}
 will also be nematic.
Our prediction can be tested by measuring  the
electronic anisotropy in such moir\'{e} devices, 
using STM, transport and other experimental means.

\begin{figure}[t]
\includegraphics[width=0.6\columnwidth]{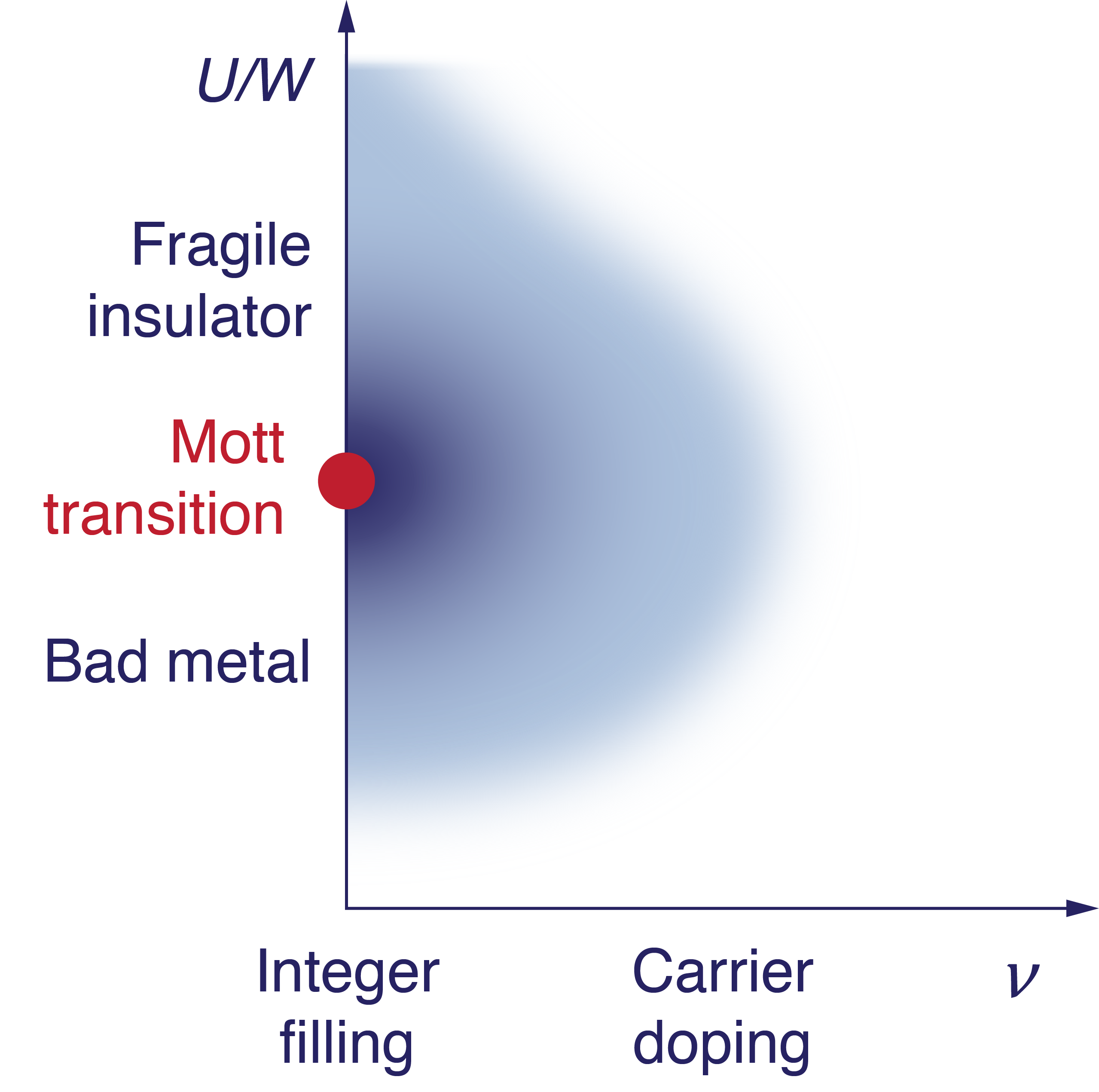}
\vskip 0.15 in
\caption{
{\bf Schematic phase diagram involving fragile insulator and bad metal.} At an integer filling, 
the
fragile insulator and bad metal are 
on
the
 two sides of an electron localization transition (red point, a Mott transition) as a function of the interaction $U/W$.
Both control the fluctuations in the magnetic and valley channels, thereby anchoring the physics in the regime where
the carrier concentration ($\nu$) is away from 
the integer filling
(the shaded blue region).
}
\label{fragileinsulator}
\end{figure}

Our finding also points towards a new organizing principle for the overall phase diagram of the 
narrow band moir\'{e} systems, as illustrated in Fig.\,\ref{fragileinsulator}.
Our calculations show that the electronic nematic order appears both in the fragile insulator and in the bad metal.
This illustrates the insensitivity of the underlying correlation physics to whether the parent 
system happens to be placed on either side of the Mott transition,
suggesting that both the fragile insulator and bad metal
can anchor the overall phase diagram.
The
emerging picture is that the system away from any integer filling can be considered as being 
controlled by the Mott transition, the 
electronic localization-delocalization transition at the integer filling that links the fragile insulating and bad metallic regimes.
By extension, when the carrier concentration is tuned away from half filling, the physics in 
the two
cases is expected to be similar.

This picture is important  for understanding another puzzle that 
has been highlighted by very recent experiments.  
It follows from Fig.\,\ref{fragileinsulator} that, when the
correlation strength is tuned down and the fragile insulator yields to a 
bad metal, the system at half filling continues to anchor qualitatively similar electronic behavior in the overall
phase diagram, including the emergence of superconductivity.  
This is precisely what have been observed by the recent experiments 
of Refs.\,\citenum{Stepanov2020Metallic,Saito2020Metallic,Arora2020Metallic}. By extension,
these  experimental
observations are fully compatible with the superconductivity being (primarily) driven by electron-electron interactions.

The overall picture, Fig.\,\ref{fragileinsulator}, also suggests that  the physics of the integer-filled moir\'{e} systems
adiabatically evolves when the normalized interaction
$U/W$ is further enhanced from the fragile insulator/bad metal regime,
where it is of order unity, to the regime where $U/W$ is even larger and
 the correlated insulator is no longer fragile. The latter is likely the case in the recently realized moir\'{e} systems
based on the transition-metal dichalcogenides\cite{Tang2020,Regan2020,Wang2020}.
In that regime, a robust Mott insulator at the integer fillings 
is expected to anchor the correlated electron physics at carrier concentrations away from those fillings.
Thus, the correlation physics in these systems will be adiabatically connected to those in the intermediate correlation regime,
although their energy scales, such as the exchange interactions as measured by their kinetic energy, will be smaller. 

We close with two additional observations. First, our calculation at zero temperature sets the stage for addressing
how the electronic orders melt away with increasing temperature. Because the nematic order is a composite of the 
spin degrees of freedom, it can occur even when the antiferromagnetic order parameter is fluctuating \cite{Si16.1}.
The latter corresponds to a nematic order that is not accompanied by any magnetic order.
It will be instructive to
experimentally study the temperature dependences of both the nematic correlations and magnetic responses 
in the correlated moir\'{e} systems. Second, by isolating the correlation effect in a model system, our work sets the stage to address how the interplay 
between the intermediate to strong correlations and bandstructure topology influences the
fragile insulator, bad metal
and electronic
 nematicity. The interplay promises to create new phases in the overall phase diagram, but whether 
and how it will enrich the relationship between the fragile insulator and bad metal on the one hand, and superconductivity
on the other, is an exciting open question.  Empirically, the continued emergence of new members in the family of
correlated moir\'{e} systems allows for ascertaining the similarities and differences 
between these members with differing bandstructure, which will surely illuminate this outstanding issue.

In summary, we have demonstrated an emergent fragile insulator in a graphene moir\'{e} system
for the physically relevant intermediate correlation regime. This correlated insulator is accompanied by an electronic 
nematic order, which provides a natural understanding of the 
electronic anisotropy that has been observed in
 the magic-angle twisted bilayer graphene. 
Our work thus highlights the kind of clues that the electronic nematicity
provides for the microscopic correlation physics, and motivates its search in
related moir\'{e} systems. Finally, our work reveals that the same correlation physics is anchored by the 
parent system at an integer filling of the moir\'{e} lattice regardless of whether it is a fragile insulator or a bad metal.
This finding explicates a striking puzzle 
on the phase diagram of the correlated moir\'{e} systems
that has emerged from several very recent  experiments.
As such, our work uncovers a new organizing principle for
the overall phase diagram of the correlated moir\'{e} systems, which also underscores 
the distinctive correlation parameter regime that these systems occupy compared to that for
the cuprate superconductors. The new regime of parent system revealed here is sufficiently general that it may well control
 the physics of a variety of strongly correlated quantum structures and materials beyond 
the context of moir\'{e} systems.

\bibliography{moire}

\vspace{0.3cm}
\noindent{\bf Acknowledgments}\\
We acknowledge useful discussions with D. P. Arovas, F. Becca, D. Goldhaber-Gordon,
W.-J. Hu,  A. H. MacDonald, A. Pasupathy, T. Senthil and J. Shan.
This work has been supported in part by the 
U.S. DOE, BES under Award \# DE-SC0018197 and the Robert A.\ Welch Foundation Grant No.\ C-1411.
The majority of the computational calculations have been 
performed on the Shared University Grid at Rice funded by NSF under Grant EIA-0216467, 
a partnership between Rice University, Sun Microsystems, and Sigma Solutions, 
Inc., the Big-Data Private-Cloud Research Cyberinfrastructure MRI-award funded by NSF under Grant No. CNS-1338099 
and by Rice University, the Extreme Science and 
Engineering Discovery Environment (XSEDE) by NSF under Grant No.\ DMR160057. 
Q.S. acknowledges the
hospitality of the Aspen Center for Physics, which is  supported by the NSF
(Grant No. PHY-1607611), and the Institute for Materials Science
at Los Alamos National Laboratory.

\clearpage
\clearpage

\clearpage

\clearpage

\clearpage

\noindent{\bf\large Methods}\\
\\
{\bf The multiorbital Hubbard model}~~
The lowest energy levels of the original ABC stacked graphene (see Supplementary Information, Fig.\,\ref{abc})
can be modeled as a two-band effective model with cubic band touching at $K_+$ and $K_-$ momenta of the original 
BZ\cite{Koshino2009,Zhang2010,Kumar2011}. 
The perpendicular voltage bias
generates an energy difference, $\Delta_V$, between the top and bottom layers. 
The hBN layer provides a superlattice potential (Fig.\,\ref{qpz}a), with components at the reciprocal lattice vectors 
of the moir\'{e} lattice,
 which is the origin of 
the moir\'{e} bands. The combination of these terms lead to
 a two-orbital Hubbard model defined on a triangular lattice ({\it cf.} Fig.~\ref{qpz}b), 
 as given in Ref.\,\citenum{Zhang2019Bridging}:
\begin{equation}
\begin{aligned}
H & = H_0 + H_U + H_V \, , \\
H_0 & = \sum_{\bm{k},
\alpha \sigma
}
\epsilon_{\bm{k},\alpha}
 c_{\bm{k},\alpha \sigma}^{\dagger} c_{\bm{k},\alpha \sigma}
\, , \\
H_U & = \sum_{i} \frac{U}{2} n_{i}^{2} - J_H\sum_{i} \left( \frac{1}{4} n_{+,i} n_{-,i} + \bm{S}_{+,i} \bm{S}_{-,i} \right) 
\, ,\\
H_V & = \sum_{\left< i, j\right>} \left[ V n_{i} n_{j} -\sum_{\alpha_1\alpha_2\sigma_{1}\sigma_2} 
V_H c^{\dagger}_{i,\alpha_{1}\sigma_{1}} c_{i,\alpha_2\sigma_2} c^{\dagger}_{j,\alpha_{2}\sigma_{2}} c_{j,\alpha_{1}\sigma_1} \right] 
\, .
\end{aligned}
\label{Hubbard_2orb}
\end{equation}
Here, $c_{\bm{k}, \alpha \sigma}^{\dagger}$ creates an electron of wavevector 
$\bm{k}$, valley $\alpha= +$ or $-$ and spin $\sigma$, and 
$H_0$ describes the kinetic part, with hopping parameters 
up to the $5$th nearest neighbors ({\it cf.} Fig.~\ref{qpz}b)
that 
specify
 the band dispersion $\epsilon_{\bm{k},\alpha}$.
Additionally,
$H_U$ contains the onsite interactions: the Hubbard interaction $U$ preserves 
the spin-valley U(4) symmetry, while the inter-valley Hund's coupling $J_H$ breaks this symmetry 
down to U(1)$_c \times$U(1)$_{v}\times$SU(2)$_s$. The  density and spin operators are defined as 
$n_{\alpha, i} = \sum_{\sigma} c^{\dagger}_{i, \alpha \sigma} c_{i, \alpha \sigma}$,
$n_i = \sum_{\alpha} n_{\alpha, i}$,
and  $\bm{S}_{\alpha,i} = \frac{1}{2}\sum_{ \sigma \sigma'} c^{\dagger}_{i,\alpha\sigma} 
\tau_{\sigma\sigma'}c_{i,\alpha \sigma'}$ respectively, with $\tau$  being the Pauli matrices. 
Finally, $H_V$ contains the nearest-neighbor interactions: $V$ is for density-density, and $V_H$ for spin-valley exchange.

\noindent
{\bf Variational Monte Carlo method}~~
We follow the VMC approach of Ref.\,\citenum{Hu2019}, 
which incorporated a spin Jastrow factor 
in the Jastrow-Slater wavefunction\cite{Capello2004} (in addition to the usual density Jastrow factor)
to treat the correlation
effect of the Hund's coupling non-perturbatively. 
The $L \times L$ geometry of our simulation is illustrated in 
the Supplementary Information, Fig.\,\ref{df}.

\noindent
{\bf U(1) slave-spin method}~~
In the U(1) slave-spin method\cite{Yu2012}, the electron creation operator is 
expressed in terms of an $xy$ spin operator $S^+_{i,\alpha\sigma}$,
which represents the charge degree of freedom,
and a fermionic `spinon' operator $f_{i,\alpha\sigma}$:
$ c_{i,\alpha\sigma}^\dag = S^+_{i,\alpha\sigma} f_{i,\alpha\sigma}^\dag$.
This is accompanied by a local constraint: $S_{i,\alpha\sigma}^z+1/2=f_{i,\alpha\sigma}^\dag f_{i,\alpha\sigma}$. 
A set of self-consistent equations provide a saddle-point description, which results in 
the
quasiparticle
weight $Z_{\alpha\sigma} = |\left <PS^+_{\alpha\sigma}P\right >|^2$, 
where $P$ is a projection operator 
that enforces
the local constraint.

\noindent
{\bf Magnetic order}~~
We calculate the spin structure factor defined as
\begin{equation}
\label{m2}
S({\bf Q})=\frac{1}{N} \sum_{i,j} \,\langle {\bf S}_i \cdot {\bf S}_j \rangle \, e^{i {\bf Q}  \cdot ({\bf R}_i-{\bf R}_j)},
\end{equation}
at $\bf{Q} = \bf{M}$, where $N$ is the number of sites.
The magnetic order parameter is $m^2 = S({\bf M}) /N$.

\noindent
{\bf Nematic order}~~
The model, defined on the triangular lattice, has a $C_6$ rotational symmetry. However, weak terms that have been neglected 
in the Hamiltonian would reduce the symmetry to $C_3$. We have thus constructed the 
 possible channels of 
nematic order from the irreducible representations of both the $D_6$ and $D_3$ point groups. The result is shown in 
Table~\ref{table:symm}. 

In TBG systems, the symmetry group is $D_6$ or $D_3$ depending on the twisting center, and the majority of the spectral weights 
stays on an effective triangular moir\'{e} superlattice. The same symmetry classification of the nematic orders applies
to the TBG system. 
\vskip 0.3 cm

\begin{table}[ht]
 \renewcommand{\arraystretch}{1.5}
\centering
\begin{tabular}{ |c | c | c | }
\hline 
Irr. Rep. ($D_6$) & Irr. Rep. ($D_3$)   & Nematic order\\
\hline
$B_1$ &$A_1$& $\frac{1}{\sqrt{6}} \sum_r (-1)^r B_r$\\
\hline
$E_1 $&$E $&$\frac{1}{\sqrt{6}} \sum_r e^{i\frac{r\pi}{3} } B_r$  ,  $  \frac{1}{\sqrt{6}} \sum_r e^{i\frac{5r\pi}{3} } B_r$\\
\hline 
$E_2 $&$E$&$ \frac{1}{\sqrt{6}} \sum_r e^{i\frac{2r\pi}{3} } B_r$  ,  $ \frac{1}{\sqrt{6}} \sum_r e^{i\frac{4r\pi}{3} } B_r$\\
\hline
\end{tabular}
\vskip 0.25cm
\caption{{\bf The classification of the nematic order.}
 Here, the nearest-neighbor-bond 
 variables
 are 
$B_r =\frac{1}{N} \sum_{i} \left < {\bf S}_i \cdot {\bf S }_{i+e_r} \right >$,
where $\{e_r\}_{r=1,..,6}$ denote the set of six nearest neighbors, with $e_1, e_2, e_3$ shown in 
Fig.\,\ref{qpz}a,
and $e_4,e_5,e_6=-e_1,-e_2,-e_3$.}
\label{table:symm}
\end{table}
The nematic order that is important for the present work is in the $E_2$/$E$ channel, in the two classification
schemes respectively:
\begin{equation}\label{sigma1}
\sigma=\frac{1}{N}\sum_{i} \left [\langle {\bf S}_{i} \cdot {\bf S}_{i+e_1}\rangle + e^{i\frac{2\pi}{3}} \langle {\bf S}_{i} 
\cdot {\bf S}_{i+e_2} \rangle + e^{i\frac{4\pi}{3}} \langle {\bf S}_{i} \cdot {\bf S}_{i+e_3}\rangle \right ] \, .
\end{equation}

\noindent
{\bf Data availability}~~
The data that support the findings of this study are available from the corresponding author
upon reasonable request.

\clearpage

\setcounter{figure}{0}
\setcounter{equation}{0}
\makeatletter
\renewcommand{\thefigure}{S\@arabic\c@figure}
\renewcommand{\theequation}{S\arabic{equation}}

\begin{figure}[t]
\centering
\includegraphics[width=0.85\columnwidth]{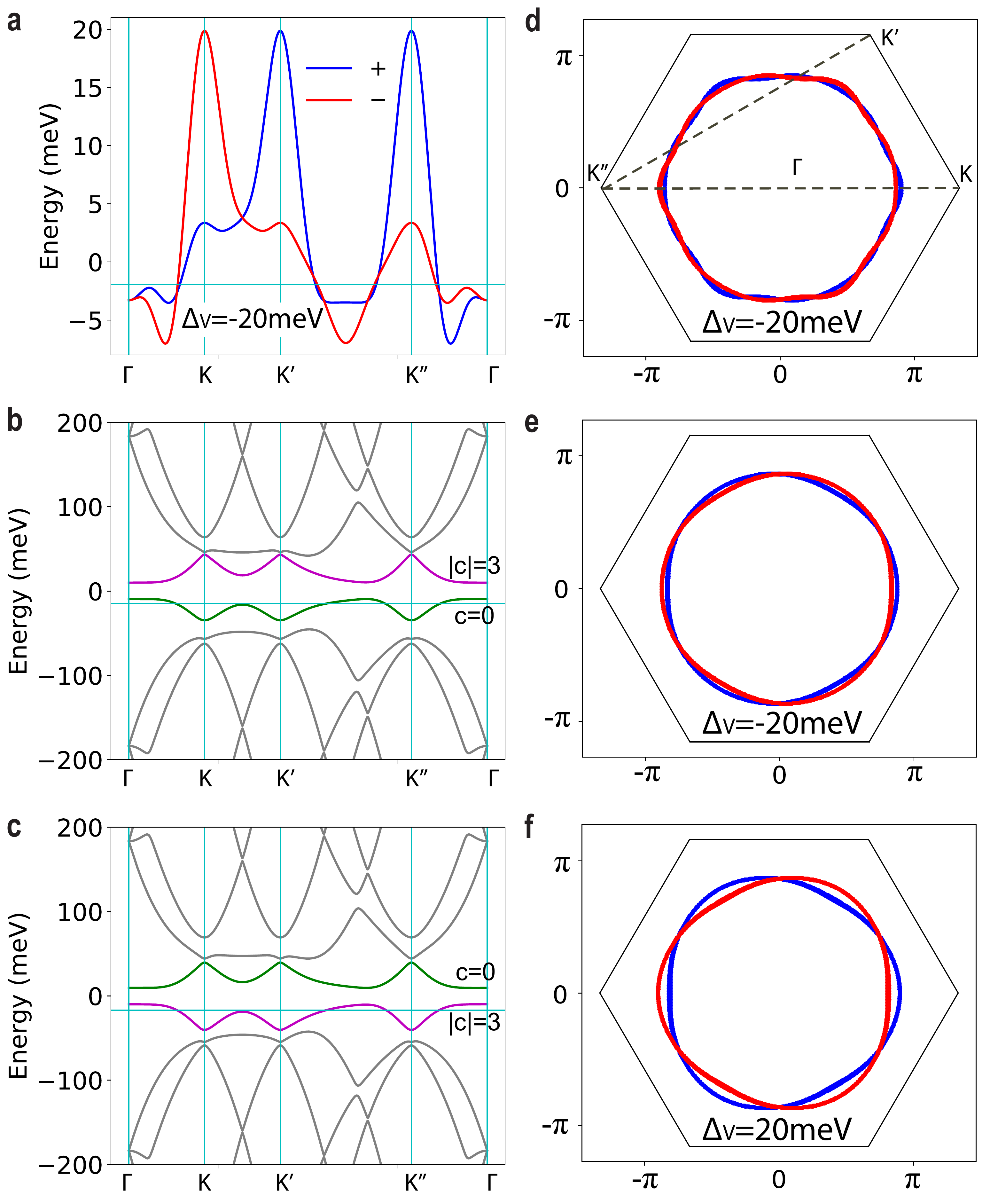}
\vskip 0.15 in
\caption{{\bf Bandstructure and
 Fermi surface.}
{\bf a,} The band structure 
of the two-orbital tight-binding model,
with the perpendicular bias energy 
$\Delta_V=-20$ meV,
for both valleys. 
The horizontal cyanic line represents the Fermi energy for half filling. {\bf b}, 
The band structure of the continuum model 
for the $+$ valley,
with $\Delta_V=-20$ meV.
The Fermi energy crosses the band with the Chern number $c=0$. {\bf c,} The counterpart of {\bf b} for 
$\Delta_V=20$ meV. The Fermi energy crosses the 
band
with nonzero $c$.
 {\bf d,\,e,\,f,}
The Fermi surface at half filling for the tight binding model,
and for the continuum models with
$\Delta_V=-20$ meV and $\Delta_V=20$ meV.
}
\label{bs}
\end{figure}

\noindent{\bf\Large Supplementary Information}\\
\\
\noindent {\bf Bandstructure and Fermi surface}\\
We outline the bandstructure, both over an extended energy range and for the bands retained in the model, 
and the Fermi surface\cite{Chen2019-1,Zhang2019Nearly,Zhang2019Bridging}.
We will use the notation of Ref.\,\citenum{Zhang2019Bridging} for the most part.
If one firstly ignores the hBN layer and focuses on the ABC stacked trilayer graphene ({\it cf.} Fig.~\ref{abc}),
the bare hopping parameters are known in the literature\cite{McCann2013}, and there is an energy difference 
$\Delta_V$ between the top and bottom layers.
There is 
a
cubic band touching at each of the two momenta, $K_{lbz}$ and $K'_{lbz}$ of the original (large)
BZ,
which are labeled as valley
 $+$ and $-$.
One can integrate out the higher energy states to construct an effective Hamiltonian for the electron states associated 
with the top (t) and bottom (b) layers. The low energy behavior in each valley is described by a two band model, 
counting the contributions from the $A$ sublattice of the top layer and $B$ sublattice from the bottom layer~\cite{Koshino2009}.
The other degrees of freedom are gapped out because of the
large direct interlayer hybridization:
{\it cf.} the tight-binding parameters, as illustrated in Fig.~\ref{abc}, are
$\gamma_0 \approx -3$ eV, $\gamma_1 \approx 380$ meV, $\gamma_3 \approx  293$ meV, $\gamma_4 \approx 144$ meV 
(Ref.\,\citenum{McCann2013}).
The aligned hBN substrate creates a potential for the adjacent graphene layer. This potential comprises
components at the mori\'e superlattice Bravais vectors for each of the two valleys.
Diagonalizing this Hamiltonian numerically up to the 5th $q_M=\frac{4\pi}{3a_M}$, where $a_M \approx \frac{a_1a_2}{a1-a2}\approx58a$
with $a_{1,2}$ being the lattice constants of the TLG and hBN, yields the band dispersion in the extended energy range
(Fig.\,\ref{bs}b,c). The two sets of moir\'{e} bands are separated by a gap, as opposed to be gapless with Dirac points in the 
TBG case.

For different signs 
of $\Delta_V$, the Fermi energy crosses two different sets of the moir\'{e} bands: 
Those
 for $\Delta_V>0$ have nonzero Chern numbers,
while those for $\Delta_V<0$ do not. In the latter case, the effective Hamiltonian is a two-band Hubbard model 
defined on the triangular 
lattice illustrated in Fig.\,\ref{qpz}a, as presented in Eq.\,(\ref{Hubbard_2orb}) (Ref.\,\citenum{Zhang2019Bridging}).
The kinetic part has the following form:
\begin{equation}
\begin{aligned}
H_0 = -\sum_{ij} t_{ij} c_{+\sigma}^{\dagger} c_{+\sigma} 
-\sum_{ij} t_{ij}^{*} c_{-\sigma}^{\dagger} c_{-\sigma} + h.c.,
\end{aligned}
\end{equation}
where $\pm$ is the valley index and $\sigma=\uparrow,\downarrow$ is the spin index. 
The time reversal symmetry dictates $\epsilon_{\bm{k},+} =\epsilon_{-\bm{k},-}$.
The complex hopping terms break the spin-valley U(4) symmetry down to U(2)$_{+}\times $U(2$)_{-}$ (Refs.\,\citenum{Zhang2019Nearly,Zhang2019Bridging}). 
The tight-binding parameters, as illustrated in Fig.\,\ref{qpz}b of the main text;  for the case 
of $\Delta_V=-20$ meV are
$t_{1} = 1.583e^{i0.169\pi}$ meV, $t_{2} = -1.108$ meV, $t_3=0.732e^{-i0.653\pi}$ meV 
and $t_{4} = t_{5}^{*} = 0.323e^{-i0.069\pi}$ meV
(Ref.\,\citenum{Zhang2019Bridging}). Those for the other symmetry-related bonds are 
 generated by the $C_6$ rotation and $M_y$ reflection. This mori\'e band structure
 is shown in Fig.\,\ref{bs}a, with a bandwidth $W=26.9$ meV. 
The path 
within the moir\'{e} BZ,
along which the band structure is shown,
can be found
 in Fig.\,\ref{bs}d. 
The 
corresponding
Fermi surfaces 
at half filling
are presented in Fig.~\ref{bs}d,e,f.

\begin{figure}[t]
\centering
\includegraphics[width=0.9\columnwidth]{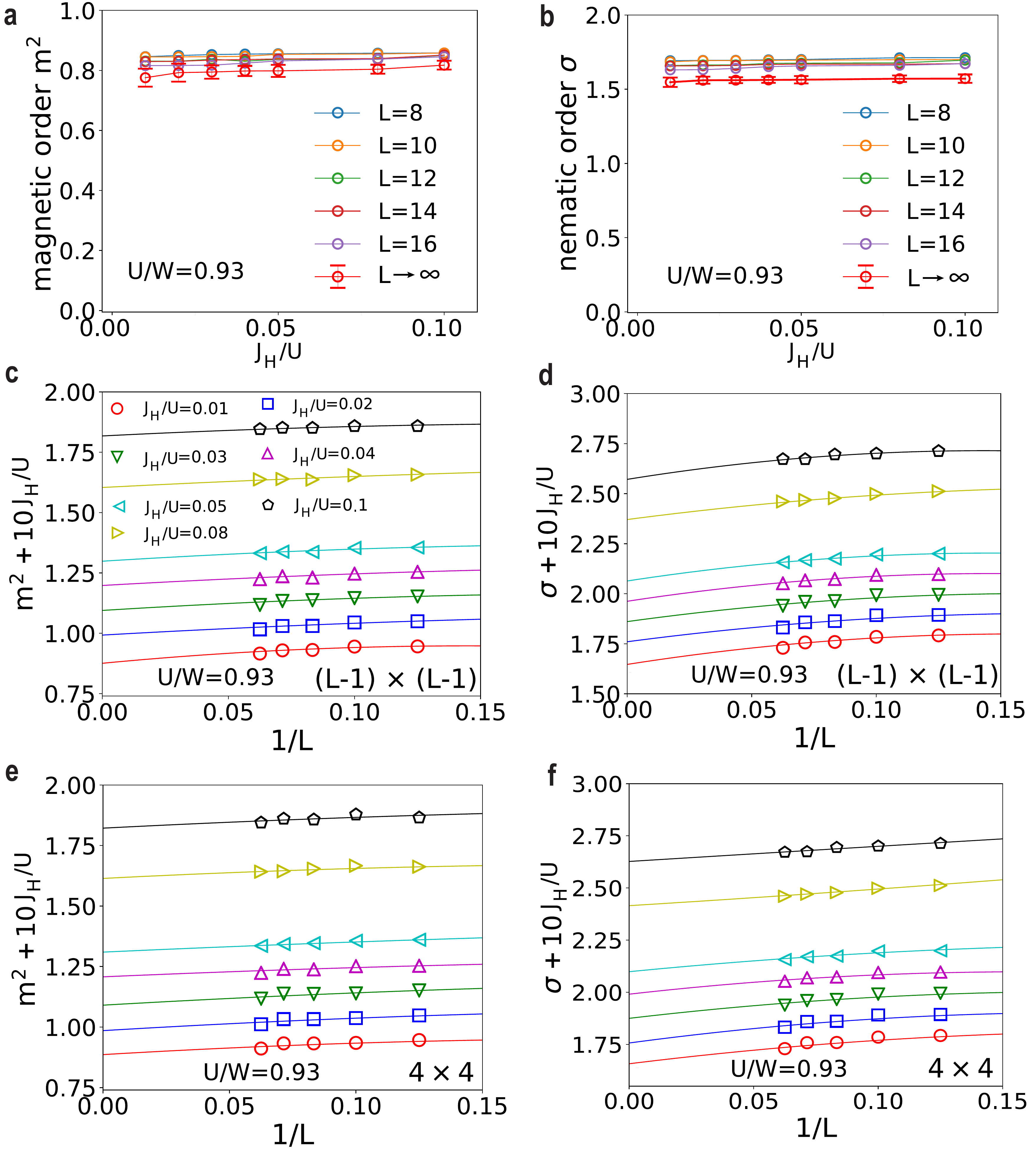}
\vskip 0.15 in
\caption{
{\bf Size dependence and finite size scaling of the 
order parameters.
}
{\bf a,} Magnetic order parameter $m^2$ as a function of $J_H/U$ for
a
 fixed $U/W$, at different sizes $L=8,10,12,14,16$ and 
in the limit $L \rightarrow \infty$ based on finite size scaling. {\bf b,} 
The counterpart
of {\bf a} for the nematic order parameter $\sigma$.
{\bf c,\,d,} Finite size scaling of the magnetic and nematic order parameters, 
measured in the interior
$(L-1) \times (L-1)$ region,
for the different values of $J_H/U$ 
at a
 fixed $U/W$.
The different $J_H/U$ cases are shifted by ``$10J_H/U$" for clarity.
{\bf e,\,f,} The counterparts of {\bf c} and {\bf d} with the order parameters measured in the central
$4 \times 4$ region.}
\label{ord}
\end{figure}

\noindent {\bf Details of the Variational Monte Carlo method}\\
The VMC approach is adapted from that of Ref.\,\citenum{Hu2019}, 
which considered a square lattice. 
In the present study, our
  model is defined
on a triangular lattice. 
As in Ref.\,\citenum{Hu2019}, a spin Jastrow factor is used, in addition to the usual density Jastrow factor,
to treat the correlation effect
of the Hund's coupling.
The uncorrelated state $|\Phi_0\rangle$ is specified by the following 
auxiliary (quadratic) Hamiltonian~\cite{Tocchio2016,Franco2018}:
\begin{equation}\label{eq:aux}
\begin{aligned}
\mathcal{H}_{\mathrm{aux}} = & - \sum_{ij, \alpha \sigma} (1+\delta \tilde{\alpha}_{ij}) t_{ij} \left( c^{\dagger}_{i, \alpha\sigma} c_{j,\alpha\sigma} +h.c. \right) 
 -\sum_{ij,\sigma} \delta \tilde{t}_{ij} \left( c^{\dagger}_{i,+\sigma}c_{j,-\sigma} + h.c. \right)
 \\
&
 + \sum_{i,\alpha\sigma} \tilde{\mu}_{\alpha} c^{\dagger}_{i,\alpha\sigma} c_{i,\alpha\sigma}
 + 
 \Delta_{\alpha}^{AFM} \left( 
 \sum_{i,\alpha} e^{i{\bf Q} _{mag} \cdot {\bf R}_i} 
 c_{i,\alpha\uparrow}^{\dagger} c_{i,\alpha\downarrow} + 
  h.c.
 \right) \\
& +\sum_{i,\alpha \sigma}  \alpha \Delta^{AFV} \left( e^{i {\bf Q}_{v} \cdot {\bf R}_i}  c_{i,\alpha\sigma}^{\dagger} c_{i,\alpha\sigma} + h.c. \right ) \\
\end{aligned}
\end{equation}
where $\delta\tilde{\alpha}_{ij}$, $\delta\tilde{t}_{ij}$, $\tilde{\mu}_{\alpha}$, and $\Delta_{\alpha}^{AFM}$, 
$\Delta^{AFV}$ are variational parameters
and real. The first two terms came from the renormalization of the hopping. 
For the intra-valley hopping, we fix the phase to be the same as for the non-interacting limit, 
while introducing the amplitude scaling variables $(1+\delta\tilde{\alpha}_{ij})$. 
The presence of $\Delta^{AFM}_{\alpha}$($\Delta^{AFV}$) $\neq0$, implies magnetic (valley) order.
By choosing ${\bf Q}_{mag}$ (${\bf Q}_{v}$) to be  ${\bf K}$ $(\frac{4}{3}\pi,0)$ or ${\bf M}$ $(\pi,\frac{1}{\sqrt{3}}\pi)$, 
we can have either three sublattice ($120^{\circ}$)-like or two sublattice magnetic (valley) orders. 
The geometry is shown in~\ref{df}, with the periodic boundary condition (PBC) for both directions.
Each direction has a linear dimension $L$; the total number of sites is $N=L\times L$.
For each $L$, all the results for the order parameters,
from Fig.\,\ref{en}
through Fig.\,\ref{ord}a,b,
are determined by measuring  in the interior $(L-1) \times (L-1)$ region.
The results are not sensitive to the boundary condition. We have also done measurements 
for the central $(L-2) \times (L-2)$ region and the central $4 \times 4$ region, 
and the results are similar to those from the $(L-1) \times (L-1)$ measurement.
This is illustrated by comparing
Fig.\,\ref{ord}c,d and 
Fig.\,\ref{ord}e,f.

\noindent {\bf Finite size scaling for the order parameters}\\
The magnetic and nematic orders with $L=8,10,12,14,16$ are shown in Fig.\,\ref{ord}a,b.
To extrapolate to the
thermodynamical
 limit,
 a
  finite size scaling over the system size 
is performed and is illustrated in 
Fig.\,\ref{ord}c,d. 
when the measurements were done from the interior 
$(L-1) \times (L-1)$ region for each $L$.
(Fig.\,\ref{ord}e,f, show the results for comparison,
when the measurements are done in the central 
$4 \times 4$ region for every $L$.)
Polynomial fittings, with the exponent up to $2$, are performed.
The error bars are estimated as the standard deviation between the simulated results and the estimated values 
of the fitted curves.

\noindent {\bf Hartree-Fock calculation}\\
We perform Hartree-Fock calculations to study the phase diagram and stability of the CAFM phase in the presence of 
the nearest-neighbor interaction terms of $H_V$. Here, except for the FM, CAFM, 
UAFV and 
paramagnetic
phases that have been studied by the VMC method, we also include the charge-ordered phase 
with wavevector ${\bf K}$ (CO-K), which is a three-sublattice order with the particle numbers being different 
in the different sublattices.
This type of order can be favored when the nearest-neighbor repulsion $V$ is sufficiently large.
Finally, with a sizable $t_2^2/|t_1|^2 \approx 0.49$, we can expect that the CAFM phase is energetically favored compared 
with any three-sublattice AFM. Our calculation of the AFM-K phase indicates that this is indeed the case.

\clearpage
\begin{figure}[t]
\centering
\includegraphics[width=0.7\columnwidth]{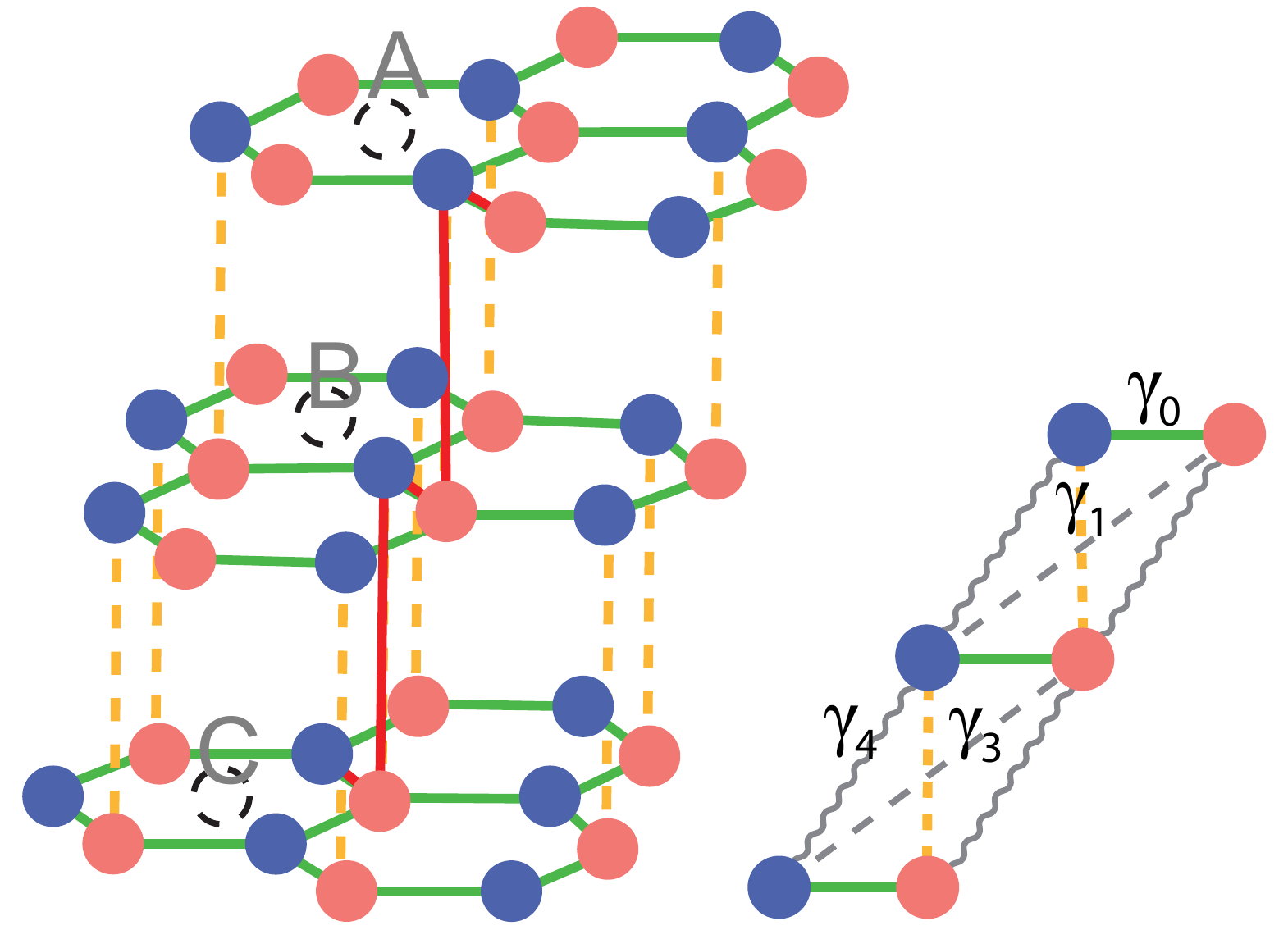}
\vskip 0.15 in
\caption{{\bf Illustration of the ABC stacked trilayer graphene.} The $\gamma$'s label the tight-binding
parameters that are used to construct the TLG
bandstructure\cite{McCann2013}.
}
\label{abc}
\end{figure}
\clearpage

\begin{figure}[t]
\centering
\includegraphics[width=0.5\columnwidth]{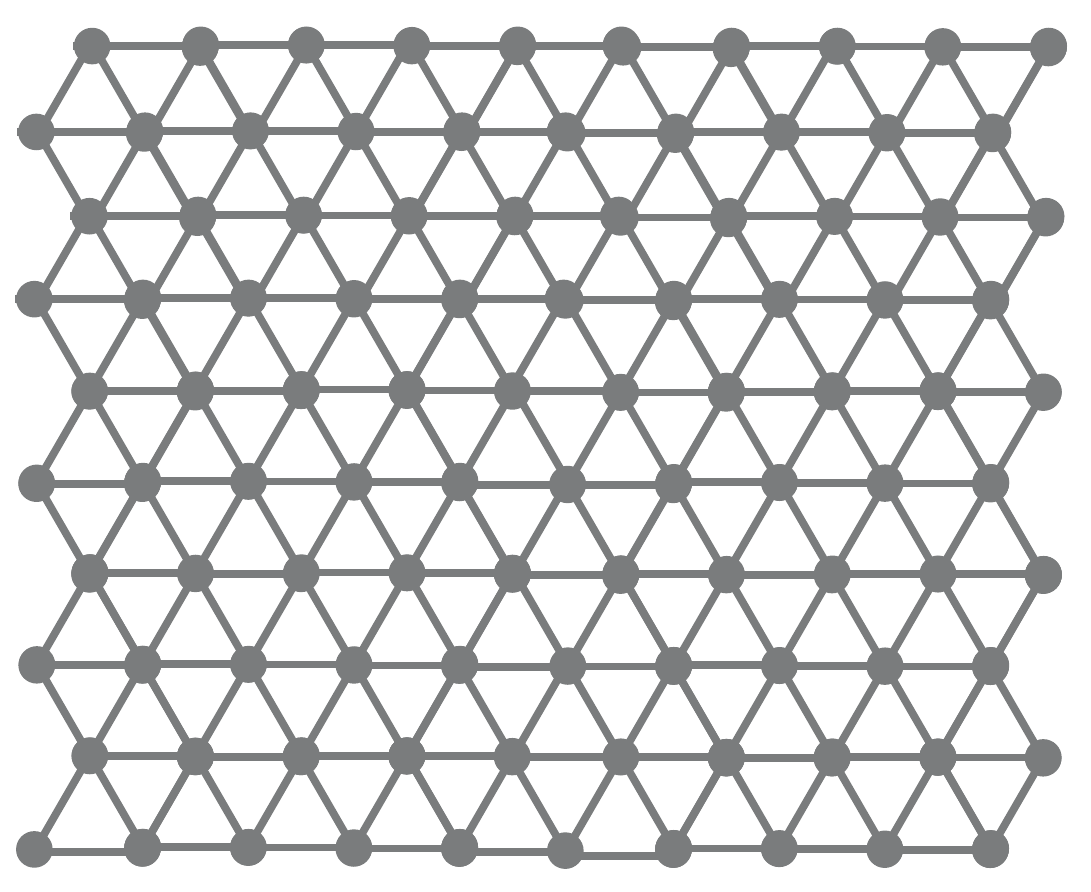}
\vskip 0.15 in
\caption{{\bf The real space structure of the mori\'e superlattice for the VMC calculation.}
Illustrated here is the $L \times L$ case  with $L=10$.
The periodic boundary condition is implemented for each direction.
}
\label{df}
\end{figure}
\clearpage

\clearpage

\clearpage

\begin{figure}[t]
\centering
\includegraphics[width=\columnwidth]{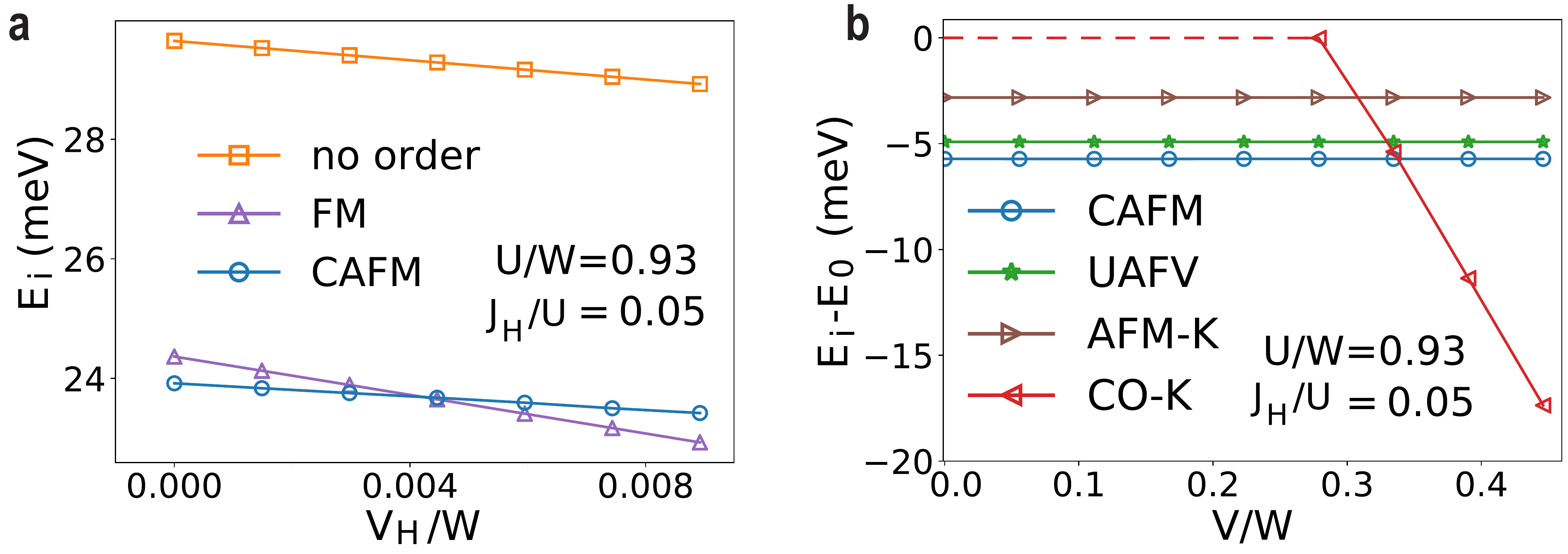}
\vskip 0.15 in
\caption{{\bf Result of the self-consistent Hartree-Fock calculation.}
{\bf a}, The ground state energies of the various symmetry-broken phases and the paramagnetic phase 
 versus the 
nearest-neighbor exchange interaction
$V_H/W$ with fixed $U/W=0.93$ and $J_H/U=0.05$. Here $i$ includes no order, ferromagnetic (FM) and collinear antiferromagnetic order 
(CAFM). {\bf b,} The difference between the ground state energy of a broken symmetry phase $i$ and that of the paramagnetic 
phase $0$ versus the nearest-neighbor density-density interaction 
$V/W$. Here, $i$ includes CAFM, the uniaxial antiferrovalley (UAFV) order, collinear 
antiferromagnetic order with wavevector ${\bf K}$ (AFM-K) and charge order with wavevector ${\bf K}$ (CO-K,
which has no solution in the parameter region corresponding to the dashed portion of the line).
}
\label{mf}
\end{figure}

\end{document}